\documentclass[a4paper,oneside,14pt,extrafontsizes]{memoir}

\usepackage{graphicx}
\usepackage[T1]{fontenc}
\usepackage[utf8]{inputenc}
\usepackage[english]{babel}
\usepackage{amsmath}
\usepackage{amssymb}
\usepackage{authblk}
\usepackage[dvipsnames]{xcolor}

\renewcommand{\bf}{\textbf}

\DeclareMathOperator{\bp}{\bf p}

\setlrmarginsandblock{0.9cm}{0.9cm}{*}
\setulmarginsandblock{2.5cm}{2.5cm}{*}
\checkandfixthelayout

\begin{document}
\title{{\bfseries Comment on the paper SciPost Phys. Core 6, 019 (2023) by João F. Melo (arXiv:2112.09119)}}
\author{Anna Radovskaya and Andrew G. Semenov}
\affil{P.N.Lebedev Physical Institute, 119991, Moscow, Russia}
\maketitle

In the paper by João F. Melo "The propagator matrix reloaded" [1] the author provides a derivation of a formalism which includes interactions in the initial conditions for non-equilibrium quantum field theory. The main statement of this paper is that one cannot ignore interactions in the initial conditions, so quantum field theories at finite temperature are not free in the infinite past. As an example, author explicitly calculate the equal-time 4-point function, checking whether or not it would be possible to get an agreement between the various approaches. The first approach is the standard one, which uses free initial conditions in the past infinity. The second one is developed by the author. This approach includes interactions in the initial conditions explicitly. After the calculations, the author claims that the first approach gives incomplete result, that's why it should not be used for finite temperature quantum field theories.

In this comment we would like to defend the first approach and point out few peculiarities in the calculations of the author. So, we claim that the free initial conditions in the past infinity can be used safely.

Let us consider the part 5 of the paper [1]. The author calculates the following 4-point average (eq. (5.1))
\begin{equation}
    \langle\hat\phi(\bp_1,t_f)\hat\phi(\bp_2,t_f)\hat\phi(\bp_3,t_f)\hat\phi(\bp_4,t_f)\rangle_\beta
    \label{ave}
\end{equation}
in the first order in coupling $\lambda$. In the paper [1] the author chooses special case 
\begin{equation}
    E_{\bp_1}=E_{\bp_2}=E_{\bp_3}=E_{\bp_4}
    \label{wrong}
\end{equation}
However, the average (\ref{ave}) is proportional to the delta-function due to momentum conservation
\begin{equation}
    \langle\hat\phi(\bp_1,t_f)\hat\phi(\bp_2,t_f)\hat\phi(\bp_3,t_f)\hat\phi(\bp_4,t_f)\rangle_\beta=F(\bp_1,\bp_2,\bp_3,\bp_4;t_f)\delta(\bp_1+\bp_2+\bp_3+\bp_4)
\end{equation}
and one should think about this expression in the sense of distributions. In means that only three momenta $\bp_{1,2,3}$ can be fixed simultaneously, whereas the last one should be $\bp_4=-\bp_1-\bp_2-\bp_3$. That's why the condition (\ref{wrong}) can not be imposed from the beginning of the calculations (as done by the author). 

Below we provide the procedure for the calculation of average (5.1) from [1] in the first approach (free initial conditions in the past infinity, 2x2 approach). We show that the correct answer can be obtained only from the two of three diagrams considered by author if one puts the initial time to $t_0\to -\infty$ properly.

Let us turn to the calculations and consider first two diagrams ((5.2) and (5.3)) which comes from real time part of the Keldysh contour. As in the paper we put $|\bp_1|=|\bp_2|=|\bp_3|$, but do not specify $|\bp_4|$. Denote $E_{\bp_1}=E_{\bp_2}=E_{\bp_3}=E$, and $E_{\bp_4}=E_{-\bp_1-\bp_2-\bp_3}=\Omega$. In this case one has
\begin{multline}
    (5.2)\equiv D_1=\frac{\lambda}{8E^3\Omega}\int\limits_{t_0}^{t_f} dt \coth^2\left(\frac{\beta E}{2}\right)\Bigg(\sin(\Omega(t-t_f))\cos^3(E(t-t_f))\coth\left(\frac{\beta E}{2}\right)\\+3\sin(E(t-t_f))\cos^2(E(t-t_f))\cos(\Omega(t-t_f))\coth\left(\frac{\beta \Omega}{2}\right)\Bigg),
\end{multline} 
\begin{multline}
    (5.3)\equiv D_2=-\frac{\lambda}{8E^3\Omega}\int\limits_{t_0}^{t_f} dt \Bigg(\cos(\Omega(t-t_f))\sin^3(E(t-t_f))\coth\left(\frac{\beta \Omega}{2}\right)\\+3\cos(E(t-t_f))\sin^2(E(t-t_f))\sin(\Omega(t-t_f))\coth\left(\frac{\beta E}{2}\right)\Bigg).
\end{multline}
Note that if we put $\Omega=E$, then we recover the expressions (5.5) and (5.6) from [1].

In the standard approach the initial time is in the past infinity, however, one can not put $t_0=-\infty$ directly in the expressions for $D_1, D_2$ because of the poor integral convergence. It is necessary to add $e^{-\epsilon(t_f-t)}$ multiplier to the $\lambda$ with infinitely small positive $\epsilon\to+0$, which describes adiabatic interaction switching. Then at $t_0\to\infty$ one has
\begin{multline}
    D_1=\frac{\lambda}{8E^3\Omega}\int\limits_{-\infty}^{0} dt e^{\epsilon t}\coth^2\left(\frac{\beta E}{2}\right)\Bigg(\sin(\Omega t)\cos^3(E t)\coth\left(\frac{\beta E}{2}\right)\\+3\sin(E t)\cos^2(Et)\cos(\Omega t)\coth\left(\frac{\beta \Omega}{2}\right)\Bigg),
\end{multline} 
\begin{multline}
    D_2=-\frac{\lambda}{8E^3\Omega}\int\limits_{-\infty}^{0} dt e^{\epsilon t} \Bigg(\cos(\Omega t)\sin^3(E t)\coth\left(\frac{\beta \Omega}{2}\right)\\+3\cos(E t)\sin^2(E t)\sin(\Omega t)\coth\left(\frac{\beta E}{2}\right)\Bigg).
\end{multline} 
After some trigonometry
\begin{equation*}
    8\sin(\Omega t)\cos^3(E t)=3\sin((\Omega+E)t)+3\sin((\Omega-E)t)+\sin((\Omega+3E)t)+\sin((\Omega-3E)t),
\end{equation*}
\begin{equation*}
    8\sin(E t)\cos^2(Et)\cos(\Omega t)=\sin((E+\Omega)t)+\sin((E-\Omega)t)+\sin((3E+\Omega)t)+\sin((3E-\Omega)t),
\end{equation*}
\begin{equation*}
    8\cos(\Omega t)\sin^3(E t)=3\sin((E+\Omega)t)+3\sin((E-\Omega)t)-\sin((3E+\Omega)t)-\sin((3E-\Omega)t),
\end{equation*}
\begin{equation*}
    8\cos(E t)\sin^2(Et)\sin(\Omega t)=\sin((\Omega+E)t)+\sin((\Omega-E)t)-\sin((\Omega+3E)t)-\sin((\Omega-3E)t)
\end{equation*}
we face with the following integral
\begin{equation}
    \int\limits_{-\infty}^{0} dt  e^{\epsilon t} \sin(\omega t)=-\frac12\left(\frac{1}{\omega+i\epsilon}+\frac{1}{\omega-i\epsilon}\right)\to -\frac{\mathcal P}{\omega}.
\end{equation} 
which should be considered in the principle value sense. As a result one has
\begin{multline}
    D_1=\frac{\lambda \coth^2\left(\frac{\beta E}{2}\right) }{64E^3\Omega}\Bigg(-\frac{3}{E+\Omega}\left(\coth\left(\frac{\beta E}{2}\right)+\coth\left(\frac{\beta \Omega}{2}\right)\right)\\+\frac{3}{E-\Omega}\left(\coth\left(\frac{\beta E}{2}\right)-\coth\left(\frac{\beta \Omega}{2}\right)\right)-\frac{1}{3E+\Omega}\left(\coth\left(\frac{\beta E}{2}\right)+3\coth\left(\frac{\beta \Omega}{2}\right)\right)
    \\+\frac{1}{3E-\Omega}\left(\coth\left(\frac{\beta E}{2}\right)-3\coth\left(\frac{\beta \Omega}{2}\right)\right)\Bigg),
\end{multline}
\begin{multline}
    D_2=-\frac{\lambda  }{64E^3\Omega}\Bigg(-\frac{3}{E+\Omega}\left(\coth\left(\frac{\beta \Omega}{2}\right)+\coth\left(\frac{\beta E}{2}\right)\right)\\-\frac{3}{E-\Omega}\left(\coth\left(\frac{\beta \Omega}{2}\right)-\coth\left(\frac{\beta E}{2}\right)\right)+\frac{1}{3E+\Omega}\left(\coth\left(\frac{\beta \Omega}{2}\right)+3\coth\left(\frac{\beta E}{2}\right)\right)
    \\+\frac{1}{3E-\Omega}\left(\coth\left(\frac{\beta \Omega}{2}\right)-3\coth\left(\frac{\beta E}{2}\right)\right)\Bigg).
\end{multline}
Now one can take the limit $\Omega=E$, which corresponds to the condition (\ref{wrong}) used in paper [1]
\begin{equation}
    D_1=\frac{\lambda \coth^2\left(\frac{\beta E}{2}\right) }{64E^4}\left(-\frac{3\beta}{2\sinh^2\left(\frac{\beta E}{2}\right)}-\frac{5}{E}\coth\left(\frac{\beta E}{2}\right)\right),
\end{equation}
\begin{equation}
    D_2=-\frac{\lambda }{64E^4}\left(-\frac{3\beta}{2\sinh^2\left(\frac{\beta E}{2}\right)}-\frac{3}{E}\coth\left(\frac{\beta E}{2}\right)\right).
\end{equation}
It is easy to check that the sum of these two terms 
\begin{equation}
    D_1+D_2=-\frac{3\lambda\beta}{128E^4\sinh^4\left(\frac{\beta E}{2}\right)}-\frac{5\lambda}{64E^5}\coth^3\left(\frac{\beta E}{2}\right)+\frac{3\lambda}{64E^5}\coth\left(\frac{\beta E}{2}\right)
    \label{result}
\end{equation}
coincides  with the total result for the average (\ref{ave}) obtained in the paper [1]. Let us stress again that we did not consider any contribution from the interactions in the initial state (or imaginary part of Keldysh contour), which is given by the third diagram ((5.3) of paper [1]).  The key point of the calculations above is the following: if one put $\Omega=E$ before calculation of the integrals then the first term in (\ref{result}) proportional to $\beta$ is missed.

The another point is related to the $e^{\epsilon t}$ multiplier and $t_0\to -\infty$ limit. This multiplier may look artificial, however, in the Keldysh technique, diagrams are usually calculated in the frequency representation where poles of propagators have the proper $i\epsilon$ terms naturally. So in such a representation this kind of regularization is done automatically. That is why the author did not find clear explanation of this procedure in the literature. 

Moreover, the correctness of the description of the thermal physics within standard Keldysh technique (2x2 technique in the paper [1]) can be proved in each order of perturbative expansion with the help of fluctuation-dissipation relations and their generalization for the multipoint correlation functions. 

The main statement of the paper [1] is based on the observation that the standard approach with free initial conditions in the past infinity and approach with imaginary time addition to the Keldysh contour do not coincide according to the calculations in sections 4 and 5. On the example of 4-point function we have checked that these two approaches give exactly the same answers, that's why the result of the paper [1] should be reconsidered.  We claim that 2x2 propagator technique gives correct answer so there is no need to apply 3x3 matrix technique unless one considers special transient processes [2,3].

\vskip 3 cm

\noindent [1] João F. Melo, \textit{The propagator matrix reloaded}, SciPost Physics Core 6, 019 (2023).
\vskip 0,5 cm
\noindent [2] A. A. Radovskaya and A. G. Semenov, \textit{Semiclassical approximation meets Keldysh\-Schwinger diagrammatic technique: scalar $\varphi^4$}, The European Physical Journal C 81, 704 (2021).
\vskip 0,5 cm
\noindent [3] Petr I. Arseev, \textit{On the nonequilibrium diagram technique: derivation, some features, and applications}, Physics Uspekhi 58, 1159 (2015).

\end{document}